\documentclass[12pt,a4paper]{article}
\usepackage{graphics}

\newcommand\ee{\mathrm{e^+e^-}}
\newcommand\as{\alpha_{\mathrm{s}}}
\newcommand\asb{\bar{\alpha}_{\mathrm{s}}}
\newcommand\kperp{k_\perp}
\newcommand\ycut{y_{\mathrm{cut}}}
\newcommand\CF{C_\sss{F}}
\newcommand\CA{C_\sss{A}}
\newcommand\sss[1]{{\scriptscriptstyle{\mathrm{#1}}}}
\newcommand\M{{\cal{M}}}
\newcommand\N{{\cal{N}}}
\renewcommand\O{{\cal{O}}} 
\newcommand\beq{\begin{equation}}
\newcommand\eeq{\end{equation}}
\newcommand\beqn{\begin{eqnarray}}
\newcommand\eeqn{\end{eqnarray}}
\newcommand\smfrac[2]{{\textstyle{\frac{#1}{#2}}}}
\newcommand\inter{_{\mathrm{inter}}}
\newcommand\intra{_{\mathrm{intra}}}
\newcommand\bp{^{\scriptscriptstyle{(\phantom{'})}}%
\makebox[0pt]{$\hspace{-0.75em}{}'$}}

\begin{document}

\begin{titlepage}

\begin{flushright}
CERN--TH/95--225\\
hep-ph/9603281
\end{flushright}

\vspace{\fill}\vspace{\fill}

\centerline{\Large\bf The Subjet Multiplicity in Quark}
\centerline{\Large\bf and Gluon Jets}
\vspace{2ex}
\centerline{\large\bf Michael H. Seymour}
\centerline{Division TH, CERN}
\centerline{CH-1211 Geneva 23}

\vspace{\fill}

\centerline{\large\bf Abstract}
\begin{quote}
We calculate the mean number of subjets in quark and gluon jets in the
final state of $\ee$ annihilation.  Since ``quark'' and ``gluon'' jets
are scheme-dependent objects, we stress the importance of using the same
definition as in experimental analyses.  We define the jets using the
$\kperp$ algorithm at a coarse scale $y_1,$ and the subjets using a
finer scale $y_0,$ in the same algorithm.  Gluon jets are anti-tagged by
the presence of heavy quarks in both other jets.
Our result is exact to leading order in $\as,$ and resums leading and
next-to-leading logarithmic terms in the ratio $y_1/y_0$ to all orders
in $\as$.
\end{quote}

\vspace{\fill}\vspace{\fill}

\begin{flushleft}
CERN--TH/95--225\\
March 1996
\end{flushleft}

\end{titlepage}

\section{Introduction}
Leading logarithmic QCD predicts that the multiplicity of hadrons in a
gluon jet should be larger than that in a quark jet by the ratio of
their colour charges, $\CA/\CF=9/4$.  Thus one would na\"\i vely expect
the multiplicity in three-jet $\ee$ events to be $(2\CF+\CA)/2\CF=17/8$
times higher than in two-jet events.  Experimentally, however this ratio
is close to unity\cite{LEP1}.  The sub-leading corrections of relative
size $\surd\as$ and even $\as$\cite{Al}, are known for the case of jets
produced by a colour singlet source.  Although they reduce this ratio,
they are small at the energies of modern experiments, and certainly not
capable of bringing it down to unity.  When the jets are produced in a
three-jet ensemble, rather than by a colour singlet, the equivalent
corrections can be much larger\cite{pQCD}, but without assumptions
about the non-perturbative hadronization process, it is impossible to
calculate their size.

Considerable light was shed on this problem by the paper of Catani {\it
  et.~al.}\cite{CDFW2}, who considered an analogous problem that can be
solved purely perturbatively, namely the multiplicity of {\em subjets\/}
in a jet.  A jet algorithm is run twice with different cutoff scales,
$Q_0$ and $Q_1,$ and the number of jets found at each, $\M_{0,1}$.
$\M_0$ is then called the number of subjets in an $\M_1$-jet event.  By
considering the region $\Lambda_\sss{QCD} \ll Q_0 \ll Q_1,$ they were
able to study the region in which logarithmic terms dominate, as in the
hadron multiplicity problem, but which is fully under perturbative
control.  They found that colour coherence during the formation of the
three-jet state was responsible for a depletion in the multiplicity in
three-jet events relative to the na\"\i ve, incoherent, expectation.
Experimental results\cite{LEP2} have shown good agreement with the
predictions of~\cite{CDFW2} for not too small $Q_0,$ although there is
clear evidence of an additional suppression in three-jet events as one
approaches the non-perturbative region.

More recently, using their excellent ability to tag heavy quark jets,
experiments at LEP have been able to measure the subjet multiplicity in
individual jets that are anti-tagged as the gluon jet in three-jet
events\cite{LEP3}.  This allows a more direct measurement of the
multiplicity ratio between gluon and quark jets in three-jet events.
Although the results were in good agreement with Monte Carlo event
generators that incorporate coherence\cite{HW,JS,AR}, it was not
possible to compare with the results of \cite{CDFW2} for two reasons.
Firstly, their method is not able to keep track of which jet the
registered subjets are part of, it simply counts the total number of
subjets.  Secondly, the experiments found that to obtain a sufficient
lever-arm in $Q_1/Q_0$ for the logarithmic behaviour to become important
before the non-perturbative behaviour, $Q_1$ had to be rather large
$\sim Q,$ where $Q$ is the total $\ee$ annihilation energy,
while~\cite{CDFW2} work in the $Q_1\ll Q$ limit.

In this paper, we correct both deficiencies by numerically integrating
an expression that exactly reproduces both the leading order expression
in $\as$ and the next-to-leading order resummation of logarithms of
$Q_1/Q_0,$ but not logarithms of $Q/Q_1$.  Our result is therefore
uniformly reliable for all $Q_1/Q_0,$ and not too small $Q_1$.  In
section~2, we give details of the event definition, which is matched as
closely as possible to the experimental one.  In section~3, we discuss
the logarithmic behaviour and define the numerical treatment, which
allows us to resum logarithms of $Q_1/Q_0,$ without losing the exact
treatment of the $Q_1\sim Q$ region.  We compare our results with the
data of~\cite{LEP3} and find good agreement where the hadronization
corrections are expected to be small.  Finally in section~4 we briefly
discuss the dependence on the energies of the individual jets within the
three-jet sample.

\section{The Event Definition}
In order to be able resum logarithms of the jet resolution scale, jets
must be defined using an algorithm such as the $\kperp$
algorithm\cite{kperp} in which the phase-space for multiple emission
factorizes in the same way as the matrix elements.  This allows jet
multiplicities to be resummed to at least next-to-leading logarithmic
accuracy to all orders in $\as$\cite{CDFW2,CDFW1}.

The algorithm defines for every pair of particles in the final state a
`closeness' measure,
\beq
  y_{ij} \equiv \frac{k_{\perp ij}^2}{Q^2} \equiv
    \frac{2\min(E_i,E_j)^2(1-\cos\theta_{ij})}{Q^2}.
\eeq
If the smallest value of $y_{ij}$ is below a cutoff $\ycut,$ the pair
$i,j$ is merged into a single pseudoparticle with momentum
\beq
  p_{ij} = p_i + p_j,
\eeq
and the process is repeated, with pseudoparticles treated on an equal
footing with particles.  If all $y_{ij}$ are above $\ycut$ the algorithm
is terminated and all remaining particles and pseudoparticles are called
jets.

To define subjets, the algorithm is first run with a cutoff $\ycut=y_0,$
and the resulting particles and pseudoparticles are called subjets.  The
algorithm is then continued with a cutoff $\ycut=y_1>y_0,$ and the
resulting particles and pseudoparticles are called jets.  During this
second stage of clustering, the fate of each subjet is kept track of,
giving the number of subjets `inside' each jet.

At present, the exact matrix elements for the final state of $\ee$
annihilation are only known up to $\O(\as^2)$\cite{ERT}.  Since the
multiplicity of subjets in an $n$-jet event only becomes non-trivial at
$\O(\as^{(n-1)}),$ we are confined to $n\le3$ in exact calculations.
Furthermore, since the two-jet final state does not contain gluon jets,
we are only interested in $n\ge3$.  Therefore, we can only calculate a
single non-trivial term, the leading order expression for the subjet
multiplicity in three-jet events.  This can be easily found by
numerically integrating the exact matrix element within the phase-space
region given by the event definition.

So far, we have fully defined the subjet structure of the event, but to
obtain the maximum amount of information from such events, we must also
define the flavour structure.  Namely which jets are quark jets and
which are gluon jets.  Beyond leading logarithmic order, the results
will depend on this definition, so we must use the same definition as in
experimental analyses.  This is done using heavy flavour tagging:  in
three-jet events in which two of the jets contain heavy quarks, the
third is called a gluon jet.  Then, using the subjet multiplicity in
three-jet events, $\M_3,$ and in gluon jets, $\M_g,$ the subjet
multiplicity in quark jets can be found by assuming that every three-jet
event consists of two quark jets and a gluon jet,
\beq
  \M_q \equiv \smfrac12 \Bigl( \M_3 - \M_g \Bigr).
\eeq
Using this definition, $\M_q$ is a physical quantity, since both $\M_3$
and $\M_g$ are.

At $\O(\as^2)$ there are two sub-processes to consider:
$\mathrm{q\bar{q}gg}$ and $\mathrm{q\bar{q}q\bp\bar{q}\bp}$.  For each
case, one must consider what flavour to call a jet consisting of a pair
of partons of given flavours.  When one of the partons is a gluon, this
is straightforward: the flavour of the jet is the flavour of the other
parton.  However if both partons are quarks or antiquarks, one must
carefully consider how the flavour-tagging would assign the jet,
recalling that a gluon jet is defined as the untagged one in an event in
which both the other jets are tagged as quark jets.

In the case of $\mathrm{q\bar{q}gg},$ the combination
$\mathrm{q+\bar{q}}$ should be called a quark jet, since neither of the
others is a quark.

In the case of the $\mathrm{q\bar{q}q\bp\bar{q}\bp}$ final state,
$\mathrm{q+\bar{q}}$ is called a gluon jet, since both the others are
quarks.  For the combinations $\mathrm{q+q\bp}$ and
$\mathrm{q+\bar{q}'},$ the outcome will depend on the exact experimental
procedure and the corrections that are applied to it.  We make the
simplifying assumption that each jet is equally likely to be tagged, so
these are shared between gluon and quark jets in the ratio 1 to 2.
These contributions make such a small contribution to the total
cross-section that varying this assumption does not affect the final
result significantly.

It is straightforward to integrate the leading order matrix element
according to these definitions, and we obtain the results shown later,
in figure~\ref{fig:aleph}.

\section{Resumming Large Logarithmic Terms}
When the ratio of cutoffs, $y_1/y_0,$ becomes large, double-logarithmic
terms appear in the perturbative expansion,
$\M\sim\as^n\log^{2n}y_1/y_0$.  Such terms must resummed to all orders
in $\as$ to give a reliable prediction.  Although this was done
in~\cite{CDFW2}, they used the simplifying limit $y_1\ll1,$ and were
not able to keep track of which jet contained which subjet.  In this
section we correct both deficiencies, by using a very similar trick to
that first used in~\cite{S} for the subjet multiplicity in hadron
collisions.

To illustrate the general method, we first work in logarithmic
approximation, valid for $y_0 \ll y_1 \ll 1$.  We use the usual notation
$L_{0,1}=-\log y_{0,1}$.  A three-jet configuration in which the gluon
makes an angle $\theta$ with the nearer quark can increase the subjet
multiplicity in five ways: gluon emission from the opposite quark, from
the nearer quark at angles larger than $\theta,$ angles smaller than
$\theta,$ emission from the gluon, and gluon splitting to quarks.  To
next-to-leading logarithmic accuracy, one can neglect recoils in the
emission, and the multiplicity can be easily found,
\beqn
  \M_3-3 &=&
  \smfrac12 \asb (L_0-L_1) \Bigl\{ \CF (L_0+L_1-3)
     + \CF (\smfrac23L_1-1)
     + \CF (L_0+\smfrac13L_1-2)
  \nonumber\\&&\;\;
     + \CA (L_0-\smfrac13L_1-\smfrac{14}3)
     + \smfrac23 N_f \Bigr\},
\eeqn
where $\asb=\as/2\pi$.  The separate terms correspond to the different
contributions just mentioned, and add up to the expression
of~\cite{CDFW2}.

Since there is one set of terms proportional to the colour charge of a
quark, and another to the colour charge of a gluon, it is tempting to
assume that these are the subjet multiplicities in the quark and gluon
jets respectively.  However, simple kinematics shows that this is not
the case.  For example, a gluon emitted from the nearer quark at a
larger angle than $\theta$ could be merged with either the gluon or the
quark, depending on the relative azimuths.  There are non-trivial
azimuthal correlations in the soft limit of the $\mathrm{q\bar{q}gg}$
matrix element, which are not retained in the usual evolution equations
because they average to zero.  Therefore the resummed result
of~\cite{CDFW2} {\em cannot\/} predict the number of subjets that are
merged into each jet, although it can predict the total number of them.
We shall shortly see how this problem can be solved without having to
define new azimuth-dependent evolution equations.

Similarly, emission from the gluon can be merged with the quarks,
depending on the azimuth and opening angle.  However, simple kinematics
also shows that emission from any jet at angles smaller than $\theta/2$
will always be merged with that jet.  We make use of this fact by
introducing an {\em arbitrary\/} parameter $\mu$ that separates the
intrajet and interjet regions.  We consider $\mu$ to be fixed and in the
range $Q_0\le\mu\le Q_1/2$\footnote{If $Q_0>Q_1/2,$ there are no large
  logarithms to resum, so we do not use the following procedure.  In
  that case the leading order matrix element is perfectly sufficient on
  its own.}, although in general one could make it a function of the
three-jet kinematics.  We then call emission from a jet of energy $E$ at
angles smaller than $\mu/E$ intrajet emission (note that $\theta$ must
be larger than $Q_1/E$), and at angles larger than $\mu/E$ interjet
emission.  This gives us multiplicities
\beq\label{Nintra}
  \N\intra-3 = \smfrac12 \asb \Bigl\{ 2\CF (L_{\mu0}^2-3L_{\mu0})
     + \CA (L_{\mu0}^2-\smfrac{11}3L_{\mu0}) + \smfrac23N_fL_{\mu0}
     \Bigr\},
\eeq
where $L_{\mu0}=\log\mu^2/Q_0^2,$ and
\beqn\label{Ninter}
  \N\inter &=&
  \smfrac12 \asb \Bigl\{
     2 \CF[L_{\mu0}(2L_1+2L_{\mu1})
         + L_{\mu1}(2L_1+ L_{\mu1}-3)]
  \nonumber\\&&\;\;
     + \CA[L_{\mu0}(\smfrac23L_1+2L_{\mu1}-1)
         + L_{\mu1}(\smfrac23L_1+ L_{\mu1}-\smfrac{14}3)] +
         \smfrac23N_fL_{\mu1} \Bigr\},
\phantom{(99)}
\eeqn
where $L_{\mu1}=\log Q_1^2/\mu^2$.  We make the distinction between
results for physical multiplicities, which we denote $\M,$ and
theoretical multiplicities, which we denote $\N$.  Note that
\beq
  \N\inter + \N\intra = \M,
\eeq
with no dependence on $\mu$.  Since $\mu$ is an entirely arbitrary
parameter separating different treatments of the same phenomena, it
should cancel in all physical quantities, at least to the accuracy to
which we are working.  However, in general it will not exactly cancel,
so varying it will give a rough indication of the size of neglected
terms, and hence the accuracy of our calculation.  In this sense, $\mu$
plays a similar r\^ole to the familiar renormalization and factorization
scales.

We turn now to the problem of how to resum the large logarithms of
$Q_1/Q_0$ to all orders in $\as,$ while keeping exact control of the
three-jet kinematics, ie.~without assuming $Q_1\ll Q$.  From equations
(\ref{Nintra}) and (\ref{Ninter}), it is clear that the only dependence
on $Q_0$ is through $L_{\mu0},$ so we aim to resum all terms of the form
$\as^nL_{\mu0}^{2n}$ (leading) and $\as^nL_{\mu0}^{2n-1}$
(next-to-leading)\footnote{We are implicitly assuming $\mu\sim Q_1$.}.
We do this separately for the intrajet and interjet components.

\subsection*{Resumming the intrajet multiplicity}
The structure seen in equation~(\ref{Nintra}) continues to all orders,
with the intrajet multiplicity contributing two logarithms of
$\mu^2/Q_0^2$ for every power of $\as$.  Therefore in resumming it, we
must keep track of next-to-leading corrections to the evolution.  By
explicitly setting up the evolution equations, one finds that the
intrajet multiplicities are identical to those in jets of the same
flavour formed from a colour singlet at scale $\mu$ resolved with a
transverse momentum cutoff $Q_0$.  Results for these quantities, which
we denote $\N_q(Q_0,\mu)$ and $\N_g(Q_0,\mu)$ for quark and gluon jets,
were given to next-to-leading logarithmic accuracy in~\cite{CDFW1}.  We
do not bother repeating the definitions here, but it is worth noting
their expansions in the threshold region (formally defined as the region
where $\as L_{\mu0}^2$ is small but $L_{\mu0}$ is still large),
\beqn
  \N_q(Q_0,\mu) &=&
    1+\asb(\mu)\left[\smfrac12\CF L_{\mu0}^2-\smfrac32\CF L_{\mu0}\right] +
    \O(\asb^2), \label{Nqapprox}\\
  \N_g(Q_0,\mu) &=&
    1+\asb(\mu)\left[\smfrac12\CA L_{\mu0}^2-\smfrac12b L_{\mu0}\right] +
    \O(\asb^2), \label{Ngapprox}
\eeqn
where $b=\frac{11}{3}\CA - \frac{2N_f}{3}$.  In terms of these
multiplicities, the total intrajet multiplicity is simply given by
\beq
  \N\intra = 2\N_q(Q_0,\mu) + \N_g(Q_0,\mu).
\eeq
Note that $\mu$ naturally arises as the scale to use for $\as$ when we
evaluate the fixed-order expression.

\subsection*{Resumming the interjet multiplicity}
Physically, there are two ways that the interjet multiplicity can be
increased beyond leading order: either by multiple emission from the
three-jet event, or by emission from the gluon emitted at first order.
Since the leading order expression is single-logarithmic in $L_{\mu0},$
raising it to any power results in terms that are negligible to
next-to-leading accuracy.  This means that multiple emission is
negligible, and to retain next-to-leading accuracy overall we only need
to keep track of leading logarithms arising from emission from the
leading order gluon.

As we already mentioned, to keep track of which jet an interjet subjet
was merged into, we need to include azimuthal correlations between
different emissions.  Thus it seems that azimuth-dependent evolution
equations are needed.  However this is not the case, because to leading
logarithmic accuracy, the additional partons inside the interjet subjet
can be considered collinear, and so are always combined together by the
jet algorithm before being combined {\it en masse\/} with one of the
main jets of the event.  Therefore we only need to calculate the
distribution of subjets between jets to leading order in $\as,$ to give
an expression to next-to-leading logarithmic accuracy to all orders in
$\as$.  This is identical to the trick we used in~\cite{S}, where we
calculated interjet terms (in that case generated by initial-state
radiation) for the number of subjets in a hadron-collision jet, despite
being unable to obtain analytical results for the kinematic
configurations that generated them.

In calculating the number of subjets in the interjet region, one
encounters integrals of the general form
\beq
  \N\inter \sim \int_{Q_0^2}^{\mu^2} d\kperp^2 \int d\Phi
  \frac{d\sigma}{d\kperp^2d\Phi} \Theta(\kperp,\Phi) \N_g(Q_0,\kperp),
\eeq
where $\kperp$ is the transverse momentum of the interjet gluon, $\Phi$
represents the rest of the phase-space, namely an energy fraction and an
azimuth, and $\Theta$ represents the (arbitrarily complicated)
phase-space constraints.  This can always be manipulated into the form
\beq
  \label{complicated}
  \N\inter = A\inter\int_{Q_0^2}^{\mu^2} \frac{d\kperp^2}{\kperp^2}
  \asb(\kperp) \N_g(Q_0,\kperp) + \asb(\mu)B\inter,
\eeq
where $A\inter$ and $B\inter$ are arbitrarily complicated
non-logarithmic functions, and we have neglected $\N_g$ in the second
term, which is justified to next-to-leading logarithmic accuracy.
Noting that the leading order expansion of $\N_g$ is just unity, we
obtain the exact result to leading order in $\as,$
\beq\label{Ninteras}
  \N\inter^{\as} = \asb(\mu)A\inter L_{\mu0} + \asb(\mu)B\inter.
\eeq
To calculate the integral retaining $\N_g,$ we use the evolution
equation to leading logarithmic order,
\beq
  \N'_g(Q_0,\mu) = \CA
  \int_{Q_0^2}^{\mu^2} \frac{d\kperp^2}{\kperp^2}\asb(\kperp)
  \N_g(Q_0,\kperp)
\eeq
where the prime denotes logarithmic differentiation with respect to the
second index\footnote{Note that this is a factor of two smaller than the
  $\N_g'$ we defined in~\cite{S}.}
\beq
  \N'_g(Q_0,Q)\equiv Q^2\frac{d}{dQ^2}\N_g(Q_0,Q).
\eeq
Using this to rewrite (\ref{complicated}), we arrive at the same result
exact to both leading order in $\as,$ and next-to-leading order in
$L_{\mu0}$ for all orders in $\as,$
\beq\label{Ninterfull}
  \N\inter = A\inter\frac1{\CA}\N'_g(Q_0,\mu) + \asb(\mu)B\inter.
\eeq

The final point concerns the evaluation of $A\inter$ and $B\inter$.  As
we have stressed, they are arbitrarily complicated in general, and not
amenable to analytic evaluation.  However, as in~\cite{S}, it is
possible to evaluate $\N\inter$ to next-to-leading logarithmic accuracy,
without knowing their explicit forms.  We note that the threshold
expansion of $\N'_g$ is
\beq
  \N'_g(Q_0,\mu) = \CA\asb(\mu)L_{\mu0} + \O(\asb^2L_{\mu0}^3),
\eeq
and that $B\inter$ is non-logarithmic, so the relation
\beq
  \asb B\inter \frac{\N'_g(Q_0,\mu)}{\CA\asb(\mu)L_{\mu0}} = \asb B\inter
\eeq
holds to the appropriate accuracy.  Comparing equations (\ref{Ninteras})
and (\ref{Ninterfull}), we arrive at the final answer for the interjet
multiplicity in three-jet events, exact to leading order in $\as$ and
resumming next-to-leading logarithms in $L_{\mu0}$ to all orders in
$\as,$
\beq
  \N\inter = \N\inter^{\as}\frac{\N'_g(Q_0,\mu)}{\CA\asb(\mu)L_{\mu0}}.
\eeq

Put simply, this means that the full result is found by numerically
integrating the exact leading order matrix element, and multiplying the
result by the parton multiplication factor,
\beq
  \frac{\N'_g(Q_0,\mu)}{\CA\asb(\mu)L_{\mu0}}.
\eeq
Explicit forms for $A\inter$ and $B\inter$ are never needed.  It is
worth stressing again, that this correctly keeps track of which jet each
subjet is merged into to all orders in $\as$, including azimuthal
correlations.  Note that the parton multiplication factor has threshold
expansion
\beq
  \frac{\N'_g(Q_0,\mu)}{\CA\asb(\mu)L_{\mu0}}
  = 1 + \smfrac16\CA\asb(\mu)L_{\mu0}^2 + \O(\asb^2),
\eeq
so rises from unity with increasing $L_{\mu0}$.

\subsection*{The full multiplicity}
In combining the results for the two regions, we have the opportunity to
make a final embellishment that avoids the need to differentiate between
the interjet and intrajet regions when numerically evaluating the
leading order matrix element.  This is because the intrajet region
gives a contribution of the logarithmic terms (\ref{Nintra}) plus a
non-logarithmic term.  Multiplying non-logarithic terms by the parton
multiplication factor gives negligible next-to-next-to-leading
logarithmic corrections.  Therefore the full expression for the
multiplicity reads
\beqn
  \label{full}
  \M &=& \asb(\mu) \Biggl( \M^{\as}
    - 2\left[\smfrac12\CF L_{\mu0}^2-\smfrac32\CF L_{\mu0}\right]
    - \left[\smfrac12\CA L_{\mu0}^2-\smfrac12b L_{\mu0}\right] \Biggr)
    \frac{\N'_g(Q_0,\mu)}{\CA\asb(\mu)L_{\mu0}}
\nonumber\\&&
    + 2\N_q(Q_0,\mu) + \N_g(Q_0,\mu),
\eeqn
where all the dependence on $\mu$ has been explicitly displayed.

As a final check of this formula, we show that it is independent of
$\mu$ to the appropriate accuracy by taking the logarithmic derivative,
\beqn
  \label{mudif}
  \mu^2\frac{d\M}{d\mu^2} &=&
    \asb(\mu) \Biggl(
    - 2\left[\CF L_{\mu0}-\smfrac32\CF\right]
    - \left[\CA L_{\mu0}-\smfrac12b\right] \Biggr)
    \frac{\N'_g(Q_0,\mu)}{\CA\asb(\mu)L_{\mu0}}
\nonumber\\&&
    + \asb(\mu) \Biggl( \M^{\as}
    - 2\left[\smfrac12\CF L_{\mu0}^2-\smfrac32\CF L_{\mu0}\right]
    - \left[\smfrac12\CA L_{\mu0}^2-\smfrac12b L_{\mu0}\right] \Biggr)
\nonumber\\&&
    \hspace{15em}\times\Biggl(
    \frac{\N''_g(Q_0,\mu)}{\CA\asb(\mu)L_{\mu0}}
    - \frac{\N'_g(Q_0,\mu)}{\CA\asb(\mu)L_{\mu0}^2}
    \Biggr)
\nonumber\\&&
    + 2\N_q'(Q_0,\mu) + \N_g'(Q_0,\mu).
\eeqn
The first and third terms cancel each other exactly to leading order in
$\as$ and, using the leading logarithmic result
\beq
  \N_q'(Q_0,\mu) = \frac{\CF}{\CA}\N_g'(Q_0,\mu),
\eeq
to next-to-leading logarithmic accuracy for all orders in $\as$.  The
two terms in the second bracket of the second term exactly cancel to
leading order in $\as,$ but at first sight it appears that there are
non-zero terms at next-to-leading logarithmic accuracy ($\as(\mu)\times
L_{\mu0}^2 \times$ a next-to-leading logarithmic function).  However,
note that $\M^{\as}$ contains $L_0^2$ terms with the same coefficients
as the $L_{\mu0}^2$ functions, so for $Q_0\ll\mu\sim Q_1$ the first
bracket is actual single logarithmic, and the whole term is zero to
next-to-leading logarithmic accuracy.

Therefore the expression~(\ref{full}) is indeed independent of $\mu,$ as
we claimed, and gives the full result exact to leading order in $\as,$
resumming leading and next-to-leading logarithms to all orders in $\as,$
and correctly keeping track of which jet the subjets get merged into.

\begin{figure}\vspace*{-1ex}
  \centerline{
    \resizebox{!}{6cm}{\includegraphics[100pt,240pt][525pt,580pt]{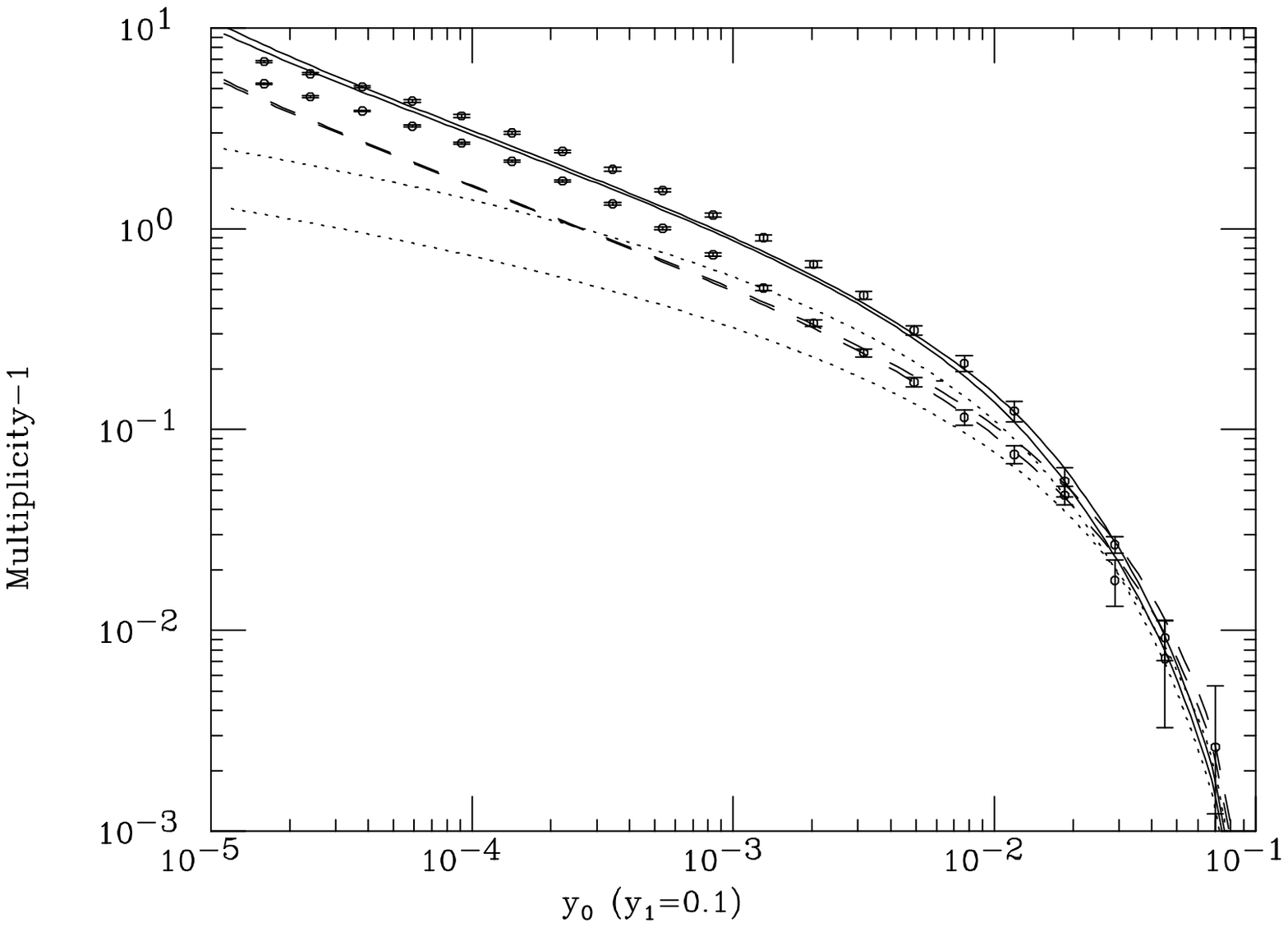}}
    \hfill
    \resizebox{!}{6cm}{\includegraphics[100pt,240pt][525pt,580pt]{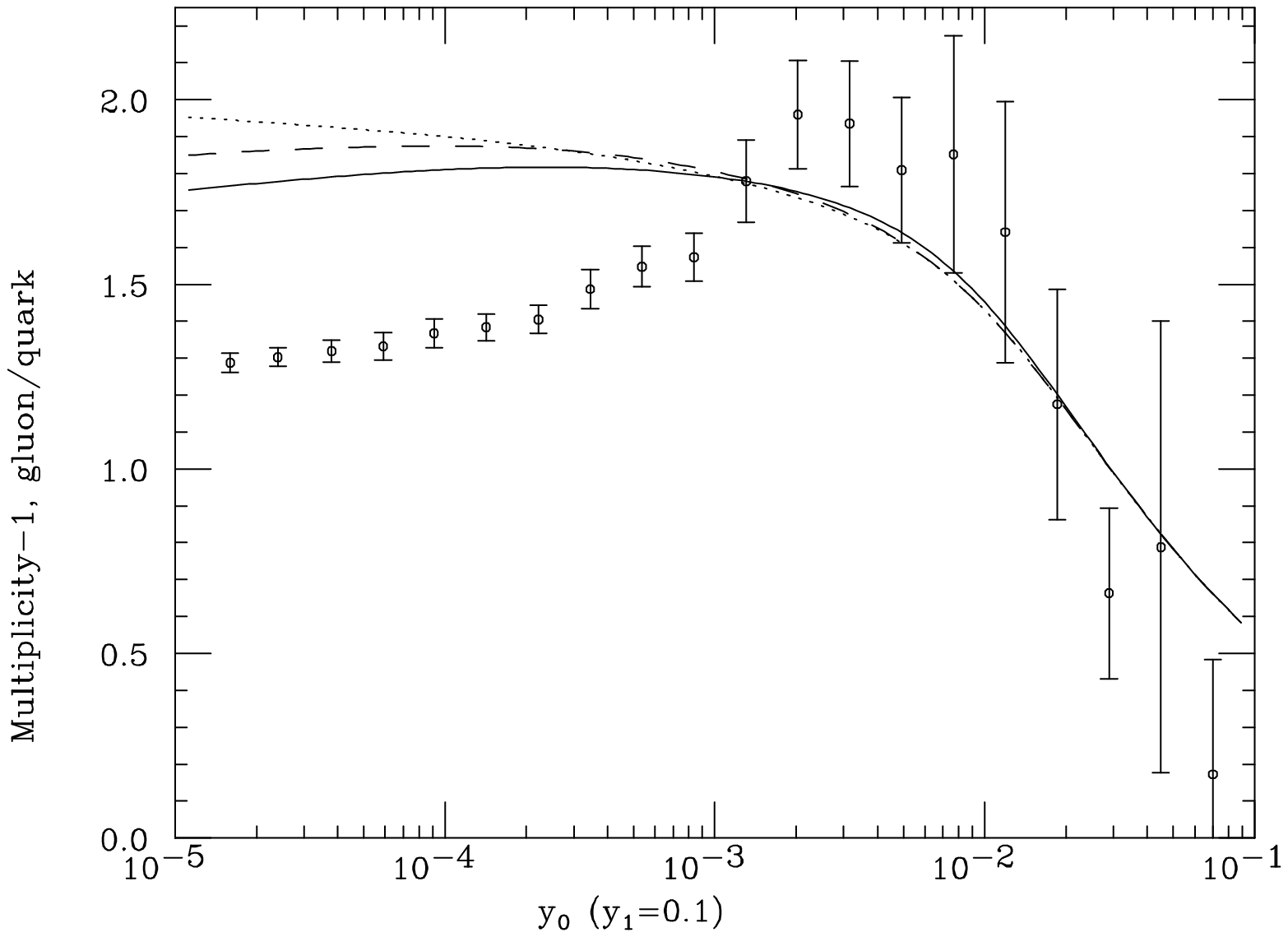}}
    }\vspace*{-2ex}
  \caption[]{{\it The subjet multiplicity in quark and gluon jets, with
      jet definition scale \mbox{$y_1=0.1$}, in comparison with ALEPH
      data}\cite{LEP3}.  (a) {\it Absolute values for gluon jets (solid)
      and quark jets (dashed) with $\mu=\sqrt{y_1}Q/2$ and
      $\mu=\sqrt{y_1}Q/4$.  Also shown are the leading order results
      (dotted)}.  (b) {\it The ratio of gluon to quark jets, with
      $\mu=\sqrt{y_1}Q/2$ (solid), $\mu=\sqrt{y_1}Q/4$ (dashed) and at
      leading order (dotted).}}
  \label{fig:aleph}
\end{figure}

In figure~\ref{fig:aleph}a we show Eq.~(\ref{full}) in comparison with
the data of~\cite{LEP3} for $y_1=0.1$.
We fix $\Lambda_{\sss{QCD}}=250$~MeV, and use the two-loop running
coupling (although this is not justified at our accuracy) corresponding
to $\as(M_{\sss{Z}})=0.120$.  We see that agreement is good down to
about $y_0=2\times10^{-3}$, which corresponds to transverse momenta of
about 4~GeV.  Below this value, the data rise above the predictions.
The barely distinguishable pairs of curves are the result of varying the
separation scale $\mu,$ with the natural choice, $Q_1/2,$ giving the
lower curve, and $Q_1/4$ giving the upper.  Also shown are the leading
order curves, which fall below the data well before the full results.

In figure~\ref{fig:aleph}b we show the ratio of the gluon and quark
multiplicities.  It can be immediately seen that in the region in which
the multiplicities rise above our predictions, $y_0<2\times10^{-3},$ the
quark multiplicity is increased more than the gluon, suppressing the
multiplicity ratio.  The effect of varying the separation scale is more
pronounced simply because the vertical axis is more compressed.  As shown
in~\cite{LEP3}, the disagreement between our results and the data can be
attributed to hadronization corrections both at small $y_0$ and the
intermediate $y_0$ values where the data lie slightly above our curve.
In fact the parton-level predictions of the JETSET program shown there
agree almost perfectly with our curves, while the hadron-level
predictions agree almost perfectly with the data.

\begin{figure}\vspace*{-1ex}
  \centerline{
    \resizebox{!}{6cm}{\includegraphics[100pt,240pt][525pt,580pt]{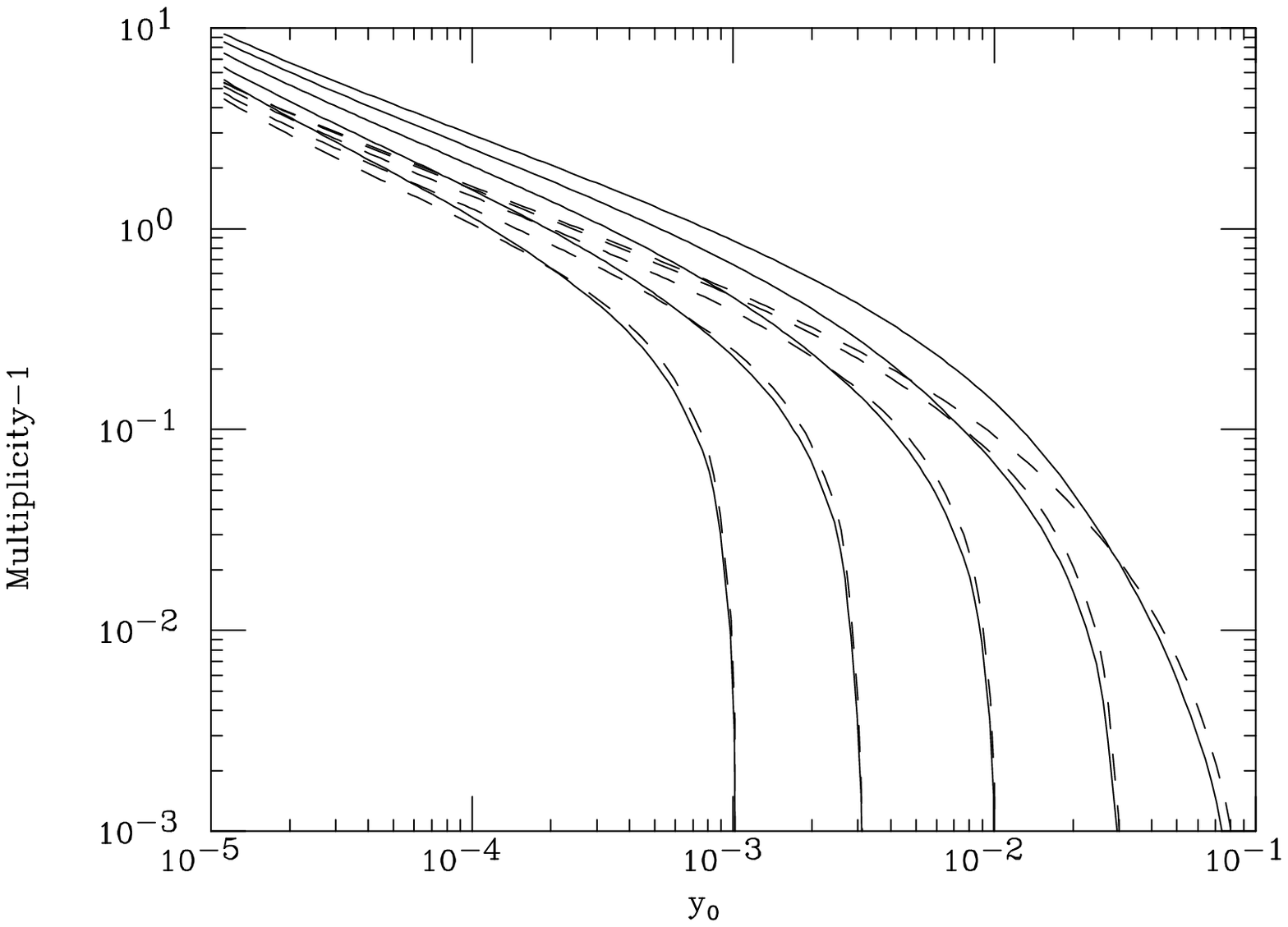}}
    \hfill
    \resizebox{!}{6cm}{\includegraphics[100pt,240pt][525pt,580pt]{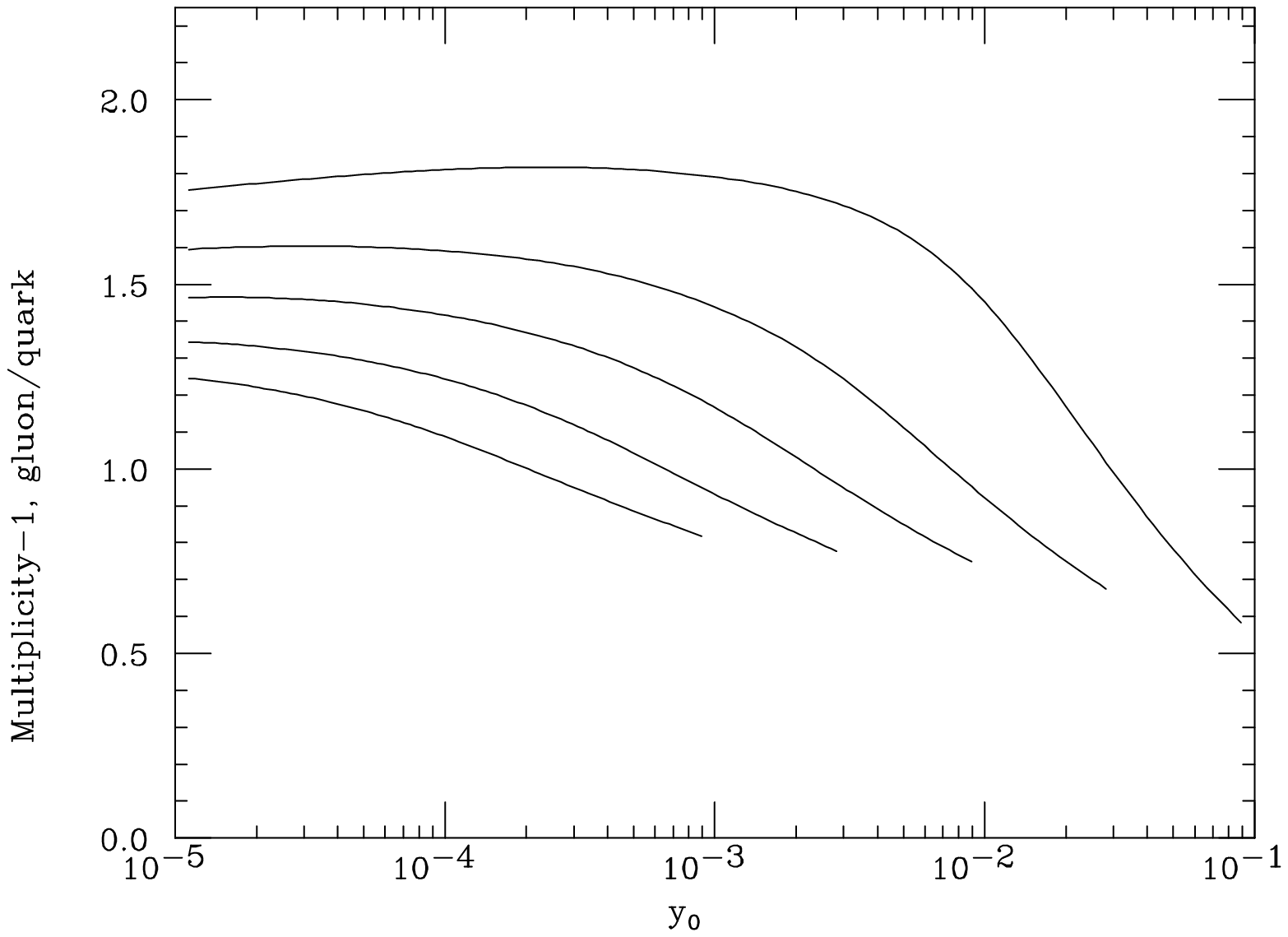}}
    }\vspace*{-2ex}
  \caption[]{{\it As in figure~\ref{fig:aleph} but for varying values of
      $y_1=$0.1, 0.03, 0.01, 0.003 and 0.001.}}
  \label{fig:y1dep}
\end{figure}

In figure~\ref{fig:y1dep} we show the $y_1$-dependence of the results,
which is considerable,
although the general features remain unchanged as $y_1$ is varied.  If a
measurement of the $y_1$-dependence could be made, it could be used to
separate the interjet and intrajet components of multiplicity, without
needing to explicitly measure the subjet directions.

\section{Jet Energy Dependence}
In the full results there is clearly a reduction in the difference
between gluon and quark jets, relative to the leading logarithmic
expectation.  One might suspect that at least part of this is because
the gluon jet tends to be the least energetic, despite the fact that the
selection $y_1=0.1$ requires all three jets to have fairly similar
energy.  This can easily be checked by binning the events according to
jet energy.

Checking Eq.~(\ref{full}) critically, one finds that it holds equally
well for the multiplicity as a function of the three-jet kinematics.
Specifically, none of the resummed components depend on the three-jet
kinematics, so if $\M^{\as}$ is binned in jet energy, then $\M$ will
also be binned correctly.

\begin{figure}\vspace*{-1ex}
  \centerline{
    \resizebox{!}{6cm}{\includegraphics[100pt,240pt][525pt,580pt]{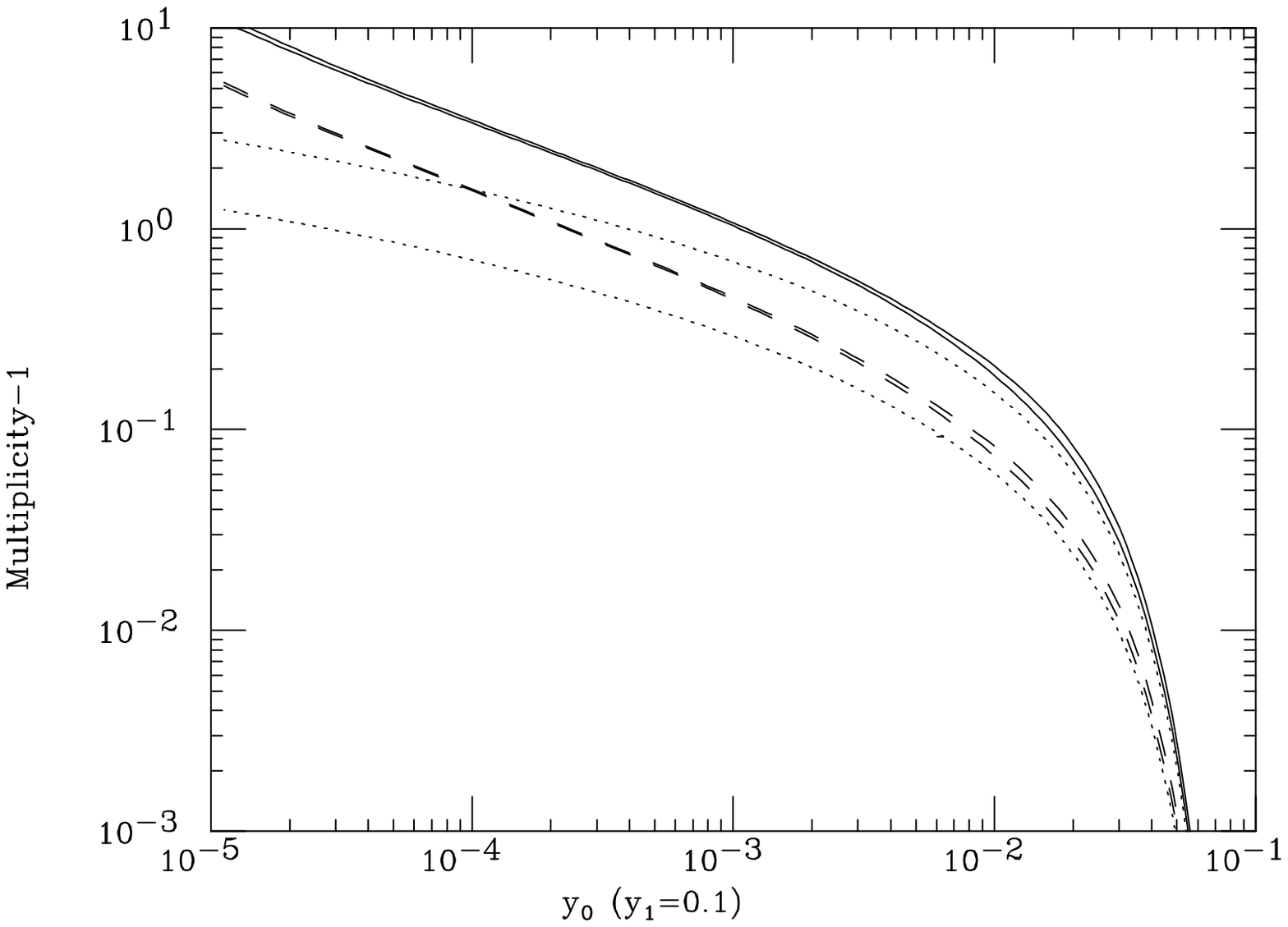}}
    \hfill
    \resizebox{!}{6cm}{\includegraphics[100pt,240pt][525pt,580pt]{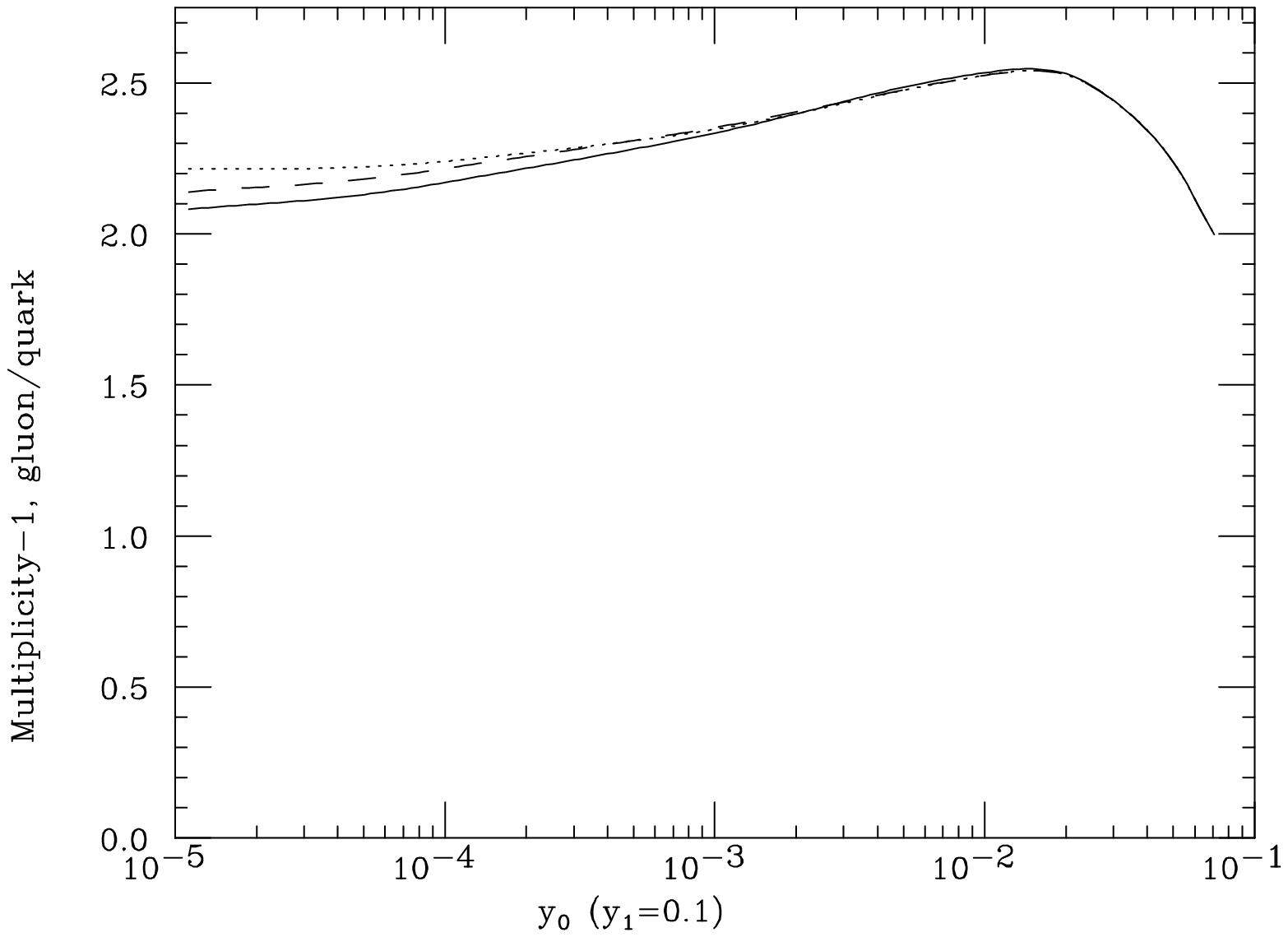}}
    }\vspace*{-2ex}
  \caption[]{{\it As in figure~\ref{fig:aleph} but for jets in the
      energy range $30<E_{\mathrm{jet}}/\mathrm{GeV}<35$.}}
  \label{fig:Edep}
\end{figure}

As an example, we show in figure~\ref{fig:Edep} the results for jets
with energy in the range $30<E_{\mathrm{jet}}/\mathrm{GeV}<35$.
We see that the multiplicity ratio is indeed enhanced, particularly at
larger $y_0$ values.

\section{Summary}
By combining numerical integration of the leading order matrix element
with analytical resummation of large logarithmic terms, we have obtained
an expression for the subjet multipicity in 3-jet $\ee$ events
(\ref{full}) that is uniformly reliable for all values of $y_0/y_1,$ and
correctly incorporates azimuthal correlations between jet and subjet
directions.  Where the hadronization corrections are estimated to be
small the results are in good agreement with data.  It is possible that
modifications to the jet definition could reduce the size of these
corrections, and the scale at which they begin to become important.  The
method used in this paper could easily be adapted to any definition in
which the systematic resummation of large logarithmic terms to all
orders is possible.  It would also be straightforward to adapt it for
any other three-jet selection or flavour-tagging method, since only the
fixed-order matrix element depends on the three-jet kinematics and this
is integrated numerically.

\end{document}